\documentclass{iopart}
\input psfig
\def\mpl{M_{\rm Pl}}
\def\half{\frac{1}{2}}

\def\etal{{\it et al.}}

\def \lleq {\lower0.9ex\hbox{ $\buildrel < \over \sim$} ~}
\def \ggeq {\lower0.9ex\hbox{ $\buildrel > \over \sim$} ~}
\def\l{\Lambda}

\def\spose#1{\hbox to 0pt{#1\hss}}
\def\simle{\mathrel{\spose{\lower 3pt\hbox{$\mathchar"218$}}
     \raise 2.0pt\hbox{$\mathchar"13C$}}}
\def\simge{\mathrel{\spose{\lower 3pt\hbox{$\mathchar"218$}}
     \raise 2.0pt\hbox{$\mathchar"13E$}}}
\def\apj{{\it Astroph.~J.~}}
\def\mn{{\it Mon.~Not. Roy.~ast. Soc.~}}

\def\aj{{\it Astron.~J.~}}
\def\prl{{\it Phys.~Rev. Lett.~}}
\def\pd{{\it Phys.~Rev.~D~}}

\def\plb {{\it Phys.~Lett.~B~}}

\def\beq{\begin{equation}}
\def\eeq{\end{equation}}
\def\ber{\begin{eqnarray}}
\def\eer{\end{eqnarray}}

\newcommand{\sq}{\lower.25ex\hbox{\large$\Box$}}

\begin{document}
\jl{6}

\title{The cosmological constant problem and quintessence.}
\author{
Varun Sahni}

\address{Inter-University Centre for Astronomy \& Astrophysics,
Pun\'e 411 007, India}
\date{\today}
\begin{abstract}
I briefly review the cosmological constant problem and the issue of
dark energy (or quintessence). 
Within the framework of quantum field
theory, the vacuum expectation value of the energy momentum tensor
formally diverges as $k^4$. A cutoff at the Planck or electroweak scale
leads to a cosmological constant which is, respectively, $10^{123}$
or $10^{55}$ times larger than the observed value, $\l/8\pi G \simeq 10^{-47}$
GeV$^4$. 
The absence of a fundamental symmetry which could set the value
of $\l$ to either zero or a very small value 
leads to {\em the cosmological constant problem}.
Most cosmological scenario's
favour a large time-dependent $\l$-term in the past (in order to generate
inflation at $z \gg 10^{10}$), 
and a small 
$\l$-term today, to account for the current acceleration of the universe
at $z \lleq 1$. Constraints arising from cosmological nucleosynthesis, 
CMB and structure
formation constrain $\l$ to be sub-dominant 
during most of the intermediate epoch $10^{10} < z < 1$.
This leads to the {\em cosmic coincidence} conundrum
which suggests that the acceleration of the universe is a recent phenomenon
and that we live during a special epoch when the density in $\l$ and
in matter are almost equal. Time varying models of dark energy 
can, to a certain extent, ameliorate the fine tuning problem (faced by $\l$),
but do not resolve the puzzle of cosmic coincidence.
I briefly review tracker models of dark energy, as well as more recent
brane inspired ideas and the issue of horizons in an accelerating universe. 
Model independent methods which reconstruct the
cosmic equation of state from supernova observations are also assessed.
Finally, a new diagnostic of dark energy -- `Statefinder', is discussed.

\end{abstract}

\bigskip

\section{Introduction}
Einstein (1917) introduced the cosmological constant $\l$ because he
believed that a closed static universe which emerged in the presence of 
both $\l$ and matter agreed with
Ernst Mach's concepts of inertia \cite{einstein,mach} 
which forbade the notion of `empty space'. 
However, the discovery by Friedmann (1922) of expanding solutions to the
Einstein field equations in the absence of $\l$, together with
the discovery by Hubble (1929) that the universe was expanding,
gave a blow to the static model\cite{friedmann,hubble}. 
Soon after, Einstein discarded
the cosmological constant.
Although abandoned by Einstein, the cosmological constant
staged several come-backs. It was soon realised that, since the 
static Einstein universe is unstable to small perturbations, one could 
construct expanding universe models which
had a  quasi-static origin in the past, thus
ameliorating the initial singularity which plagues expanding FRW models.
One could also construct models which approached the
static Einstein universe during an intermediate epoch when
the universe `loitered' with $a \simeq$ constant.
Such a model was proposed by Lemaitre (1927) and was to prove
influential later, in 1968, when it was invoked to explain an
alleged excess of quasars at a redshift $z \sim 2$.
It is also interesting that during the very same year that Einstein proposed
the cosmological constant, 
de Sitter discovered a matter-free solution to the Einstein equations
in the presence of $\l$, which had both static and dynamic representations.
The de Sitter metric was to play an important role both in connection with
steady state cosmology as well as in the construction of
inflationary models of the very early
universe.

A physical basis for the
cosmological constant had to wait until 1968, when
Ya. B. Zel'dovich puzzling over cosmological observations which
appeared to require $\l$ (the quasar excess at $z \sim 2$ alluded to earlier)
realised that one loop quantum vacuum 
fluctuations\footnote{
The presence of zero-point vacuum fluctuations was predicted by
Casimir \cite{casimir} and has been 
verified by several experiments, see \cite{cas} and references therein.}
gave rise to an energy momentum
tensor which, after being 
suitably regularised for infinities, had exactly the
same form as a cosmological constant: $\langle T_{ik}\rangle_{\rm vac}
= \l g_{ik}/8\pi G$.

Theoretical interest in $\l$ remained on the increase
during the 1970's and early
1980's with the construction of inflationary models, in which 
matter 
(in the form of a false vacuum, as vacuum polarization or as a minimally
coupled scalar-field) behaved precisely like
a weakly time-dependent $\l$-term. The current interest in $\l$
stems mainly from observations of Type Ia high redshift supernovae
which indicate that the universe is accelerating fueled perhaps by a
small cosmological $\l$-term \cite{perl99,riess98}.
\footnote{The chronology of interest in $\l$ bears a curious historical
parallel to the scientific fascination with the notion of extra
dimensions. (I thank
Nathalie Duruelle for an interesting discussion on this issue during the 
meeting.) A fourth spatial dimension was proposed by Nordstr\"{o}m (1914) and
independently by Kaluza (1921),
but the real scientific interest in higher dimensions grew 
after de Witt (1962), Kerner (1968) and others \cite{kk}
had convincingly demonstrated the deep relationship between higher 
dimensional theories on the one hand, and non-abelian gauge fields
on the other. Cosmology in a space-time with extra dimensions really
took off during the 1970's and early 1980's when grand unified and supergravity
models frequently relied on compact extra dimensions to generate the extra gauge
degrees of freedom associated with unification. Current interest in
higher dimensional cosmologies is spurred by superstring theory as well as
by `brane-world' scenario's of extra dimensions.}


\section{The cosmological constant and vacuum energy}
\label{sec:theory}

Vacuum fluctuations contributing to $\l$ generate a very large (formally
infinite) value of the cosmological constant 
$\langle T_{00}\rangle_{\rm vac} \propto \int_0^\infty \sqrt{k^2 + m^2}
k^2 dk$. The integral diverges as $k^4$ resulting in an infinite value
for $\langle T_{00}\rangle_{\rm vac}$ and hence also 
for the cosmological constant
$\l = 8\pi G\langle T_{00}\rangle_{\rm vac}$. Since each form of energy
gravitates and therefore reacts back on the space-time geometry, an infinite
value of $\l$ is expected to generate an infinitely large space-time
curvature through the semi-classical Einstein equations
$G_{00} = \frac{8\pi G}{c^4} \langle T_{00}\rangle_{\rm vac}$.
One way to avoid this is to assume that the 
Planck scale provides a natural ultra-violet cutoff to all field theoretic
processes, this results in $\langle T_{00}\rangle_{\rm vac}
\simeq c^5/G^2\hbar \sim \sim 10^{76}$GeV$^4$
which is 123 orders of magnitude larger than
the currently observed value $\rho_\l \simeq 10^{-47}$GeV$^4$.
A cutoff at the much lower QCD scale doesn't fare much better since
it generates a cosmological constant $\Lambda_{QCD}^4 \sim 10^{-3}$GeV$^4$ --
forty orders of magnitude larger than observed. Clearly the answer
to the cosmological constant issue must lie elsewhere.

The discovery of supersymmetry in the 1970's led to the hope that
the cosmological constant problem may be resolved by a judicious
balance between bosons and fermions in nature, since bosons and fermions
(of identical mass) contribute equally but with opposite sign to the
vacuum expectation value of physical quantities, so that 
\beq
\langle 0|{H_{\rm b,f}}|0\rangle \equiv \int{~dV \langle T_{00}\rangle_{\rm vac}}
= \pm \half \sum_{\bf k} \omega_{\bf k}.
\label{eq:c10}
\eeq
However supersymmetry (if it exists) is broken at the low temperatures
prevailing in the universe today and on this account one should 
expect the cosmological
constant to vanish in the early universe, but to reappear during late
times when the temperature has dropped below $T_{\rm SUSY}$.
This is clearly an undesirable scenario and almost the very opposite
of what one is looking for, since, a large value of $\l$ at an
early time is useful from the viewpoint of inflation, whereas a very
small current value of $\l$ is in agreement with observations.

\begin{figure}[tbh!]
\centerline{
\psfig{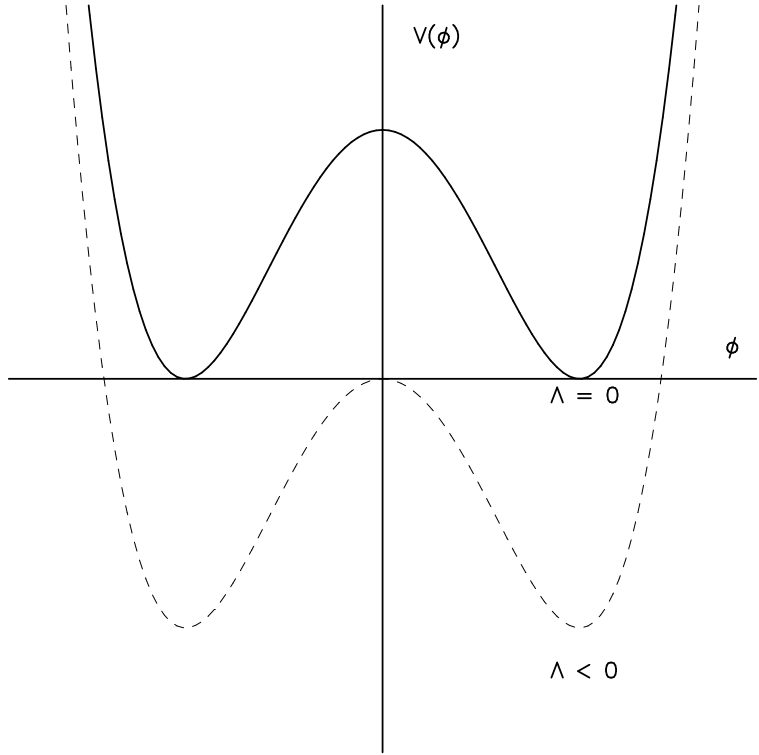}}
\caption{\footnotesize
The potential describing spontaneous symmetry
breaking has the form of a Mexican top hat. The dashed line
shows the potential before the
cosmological constant has been `renormalized' and the solid line after.}
\bigskip
\medskip
\label{fig:ssb}
\end{figure}

The cosmological constant also makes an important appearance in models with
spontaneous symmetry breaking \cite{wein89}. 
For simplicity consider the Lagrangian
\ber
{\cal L} &=& \half ~g^{ij}\partial_i\phi\partial_j\phi - V(\phi),\nonumber\\
~V(\phi) &=& V_0 -\half \mu^2\phi^2 + \frac{1}{4}\lambda\phi^4.
\label{eq:inf2a}
\eer
The symmetric state at $\phi = 0$ is unstable and the system
settles in the ground state $\phi = +\sigma$
or $\phi = -\sigma$, where $\sigma = \sqrt{\mu^2/\lambda}$, thus breaking the
reflection symmetry $\phi \leftrightarrow - \phi$ present in the Lagrangian. 
If $V_0 = 0$ then this potential results in a broken symmetry state with 
a large negative cosmological constant 
$\l_{eff} = V(\phi = \sigma) = - \mu^4/4\lambda$. In order to avoid this 
situation the value of the free parameter $V_0$ is chosen to counterbalance
$\l_{eff}$, as a result one sets $V_0 \sim \mu^4/4\lambda$ so that 
$\l_{eff}/8\pi G = V_0 - \mu^4/4\lambda \simeq 10^{-47}$GeV $^4$.
The ensuing `regularization' of the large negative cosmological
constant must be done with considerable care, since even small 
`fluctuations' in the final value of $\l$ can result in grave consequences 
for cosmology. For instance if $\l_{eff}/8\pi G < - 10^{-43}$GeV $^4$ the large
attractive force exerted by a negative cosmological constant will
ensure that the universe re-collapses before it reaches `maturity'.
The age of the universe in this case will be $<$ 
1 billion years, much too short for galaxies to form and for life
(as we know it) to emerge within the standard big bang scenario.
On the other hand if $\l_{eff}/8\pi G > 10^{-43}$GeV$^4$, the large
repulsive force generated by $\l$ will ensure that the universe 
begins accelerating
before gravitationally bound systems have a chance to form.
Such a scenario will also clearly preclude the emergence of life.

The rather small window permitted 
for life to emerge in the presence of $\l$ 
has led several researchers \cite{wein98,vilenk00,wein01}
to develop anthropic arguments for the existence of
a small cosmological constant.
A possibility which is summarised by the following sentence:
``if our big bang is just one of many big bangs, with a wide range of 
vacuum energies, then it is natural that some of these big bangs should have
a vacuum energy in the narrow range where galaxies can form, and of course it is
 just these big bangs in which there could be astronomers and
physicists wondering about the vacuum energy'' \cite{wein01}.
Anthropic arguments for $\l$
will not be examined further by me in this talk.

In the absence of a fundamental symmetry of nature which will set the
value of $\l$ to zero one has to look towards physical mechanisms
which might generate an acceptably small value of the $\l$-term today.

Exploring the connection between 
quantum fluctuations
and $\l$ Zel'dovich suggested that, after the removal of divergences,
the energy density of a virtual particle-antiparticle pair interacting
gravitationally would be \cite{zel68}
\beq
\rho_\l \sim \frac{Gm^2}{\hbar c}m~(\frac{mc}{\hbar})^3.
\label{eq:zeld}
\eeq
(This result is easy to derive if one notes that the interaction energy density
is typically $\epsilon_{vac} \equiv \rho_{vac}c^2 \sim 
\frac{Gm^2}{\lambda}/\lambda^3$ where $\lambda = \hbar/mc$ is the mean
separation between particle and antiparticle.) This possibility has not
been explored much, perhaps because the proton-antiproton (electron-positron)
contribution gives a very large (small) value for $\rho_\l$. 
Interestingly the pion-antipion mass gives just the right value
$\rho_\l = \frac{1}{(2\pi)^4}~\rho_{\rm Pl}~(m_\pi/\mpl)^6 \simeq
1.3\times 10^{-123}\rho_{\rm Pl}=6.91\times 10^{-30}~g\,cm^{-3}~.$

Purely numerological considerations also allow 
one to generate a sufficiently small
value of $\l$ through a suitable combination of fundamental constants.
For instance the fine structre constant $\alpha$ can be 
combined with the Planck density $\rho_{\rm Pl}$ to give \cite{ss99}
$\rho_\l = \frac{\rho_{\rm Pl}}{(2\pi^2)^3}e^{-2/\alpha} \simeq
1.2\times 10^{-123}\rho_{\rm Pl}=6.29\times 10^{-30}~g\, cm^{-3}~$.

A small vacuum energy may be connected to fundamental physics in
other (equally speculative) ways. It is interesting that the mass scale
associated with the scale of supersymmetry breaking
in some models,
$M_{\rm SUSY} \sim 1$ TeV,
lies midway between the Planck scale and $10^{-3}$ eV. 
The small observed value of the cosmological constant
$\rho_\l \simeq (10^{-3} eV)^4$
might therefore be associated with the vacuum in a theory which had a 
fundamental mass scale $M_X \simeq M_{\rm SUSY}^2/M_{\rm Pl}$, such that
$\rho_{\rm vac} \sim M_X^4 \sim (10^{-3} eV)^4$.

\section{A dynamical $\l$-term.}

Any fundamental theory of nature which intends to succesfully
generate $\l$ will be confronted by the
`fine tuning problem' since the currently observed value of 
the cosmological constant is miniscule when compared with either the Planck
($\rho_\l/M_{\rm Pl}^4 \sim 10^{-123}$)
or the electroweak scale
($\rho_\l/M_{\rm EW}^4 \sim 10^{-55}$). During the expansion of 
the universe the energy density in matter (radiation) decreases
as $a^{-3}$ ($a^{-4}$) while the density in $\l$
remains constant. As a result an enormous fine tuning of initial
conditions is required in order to ensure that the cosmological $\l$-term 
comes to dominate the expansion dynamics of the universe 
at precisely the current
epoch, no sooner and no later. 

The fine-tuning problem is rendered less acute if we relax the condition
$\rho_\l = $constant, and (taking the cue from Inflation)
try to construct dynamical models for
$\rho_\l$.

Phenomenological approaches to a dynamical $\l$-term
belong to three main categories \cite{ss99}:

(1) {\em Kinematic models.} 

$\rho_\l$ is simply assumed to be a function of either the cosmic time $t$
or the scale factor $a(t)$ of the FRW cosmological model.

(2) {\em Hydrodynamic models.} 

$\rho_\l$ is described by a barotropic fluid with some equation of
state $p_{\l}(\rho_{\l})$ (dissipative terms may also be present).

(3) {\em Field-theoretic models.} 
The $\l$-term is assumed to be a new physical classical field
with some phenomenological Lagrangian.

The simplest class of kinematic models
\beq
\l \equiv 8\pi G\rho_{\l} = f(a)
\eeq
is equivalent to hydrodynamic models based on an ideal fluid with
an equation of state
\beq
p_{\l}(\rho_{\l})= - \rho_{\l}(1 + {1\over 3}\, {d\ln \rho_{\l} \over d\ln a}).
\eeq
The expansion of the
universe passes through an inflection point the moment it
stops decelerating and begins to accelerate. 
If the equation of state is held constant ($w = P/\rho < -1/3$)
then the cosmological redshift
when this occurs is given by
\beq
(1 + z_{\rm a})^{-3w} = -(1 + 3w)\frac{\Omega_X}{\Omega_m}.
\eeq
We find that $z_{\rm a} \simeq 0.7$ 
for the cosmological constant ($w = -1$)
with $\Omega_\l \simeq 0.7$ and $\Omega_m \simeq 0.3$.
The acceleration of the universe is therefore a {\em very recent} phenomena.
This fact 
is related to the {\em cosmic coincidence} conundrum since it appears 
that we live during a special era 
when the density of dark matter and dark energy are comparable.
The cosmic coincidence puzzle remains in place even if we relax the 
assumption $w = -1$ and allow dark energy to be time dependent.
Indeed, it is easy to show that
the equality between dark matter and dark energy takes
place at
$(1 + z_{\rm eq})^3 = (\Omega_\l/\Omega_m)^{-1/w}$. For a cosmological constant
this gives $z_{\rm eq} \simeq 0.3$ and $z_{\rm a} > z_{\rm eq}$
implying that the universe
begins to accelerate even {\em before} it becomes 
$\l$-dominated. For $w = -2/3$ 
$z_{\rm a} = z_{\rm eq} \simeq 0.5$, while for stiffer equations of state
$z_{\rm a} < z_{\rm eq}$ ($w > -2/3$) further exacerbating the
cosmological coincidence puzzle. 

\subsection{Scalar field models of dark energy}

Although the {\em cosmic coincidence} issue remains unresolved,
the fine tuning problem facing dark energy/quintessence models
with a constant equation of state can be significantly alleviated
if we assume that the equation of state is time dependent.
An important class of models having this property are scalar fields
which couple minimally to gravity and whose energy momentum tensor
is 
\beq
\rho \equiv T_0^0 = \half\dot{\phi}^2 + V(\phi),
~P \equiv -T_\alpha^\alpha = \half\dot{\phi}^2 - V(\phi).
\eeq
A scalar field rolling down its potential
{\em slowly} 
generates a time-dependent $\l$-term since 
$P \simeq -\rho \simeq -V(\phi)$ if $\dot{\phi}^2 \ll
V(\phi)$. Potentials which satisfy
$\Gamma \equiv V''V/(V')^2 \geq 1$ have the interesting property that 
scalar fields approach a common evolutionary path from a wide range of initial
conditions \cite{track}. In these so-called `tracker' models 
the scalar field density (and its equation of state) remains close
to that of the dominant background matter during most of cosmological evolution.
A good example is provided by the exponential potential
$V(\phi) = V_0\exp{(-\sqrt{8\pi}\lambda\phi/\mpl)}$ \cite{ratra,exp1}
for which
\beq
\frac{\rho_\phi}{\rho_{B} + \rho_\phi} = \frac{3(1 + w_B)}{\lambda^2}
= {\rm constant} < 0.2,
\label{eq:exp}
\eeq
$\rho_{B}$ is the background energy density while $w_B$ is the
associated equation of state.
The lower limit $\rho_\phi/\rho_{\rm total} < 0.2$ arises because of
nucleosynthesis constraints which prevent the energy density in 
quintessence from being large initially (at $t \sim few $  sec.).
Since the ratio $\rho_\phi/\rho_{\rm total}$ remains fixed, exponential
potentials on their own cannot supply us with a means of generating
dark energy/quintessence at the present epoch. However a suitable
modification of the exponential achieves this. For instance
the class of potentials \cite{sw00}
\beq
V(\phi) = V_0[\cosh{\lambda\phi} - 1]^p,
\label{eq:cosh}
\eeq
has the property that $w_\phi \simeq w_B$ at early times
whereas $\langle w_\phi\rangle = (p - 1)/(p + 1)$ at late times.
Consequently (\ref{eq:cosh}) 
describes {\em quintessence} for $p \leq 1/2$ and pressureless 
`cold' dark matter (CDM) for $p = 1$. 

A second example of a tracker-potential is provided by 
$V(\phi) = V_0/\phi^{\alpha}$ \cite{ratra}. During tracking
the ratio of the energy density of the scalar field (quintessence) 
to that of radiation/matter gradually increases
$\rho_\phi/\rho_B \propto t^{4/(2 + \alpha)}$
while its equation of state remains marginally smaller than the
background value $w_\phi = (\alpha w_B - 2)/(\alpha + 2)$.
These properties allow the scalar field to eventually dominate the density of
the universe, giving rise to a late-time epoch of
accelerated expansion. (Current observations place the strong constrain
$\alpha \lleq 2$.)

\bigskip
\begin{table*}[tbh!]
\begin{center}
\begin{minipage}[h]{0.9\linewidth} \mbox{} \vskip 18pt
\begin{tabular}{lll}
\hline
Quintessence Potential & Reference\\\hline
& \\
$V_0\exp{(-\lambda\phi)}$ & Ratra \& Peebles (1988), Wetterich (1988), \\
& Ferreira \& Joyce (1998)\\
& \\
$m^2\phi^2, \lambda\phi^4$ &  Frieman et al (1995)\\
& \\
$V_0/\phi^\alpha, \alpha > 0$ &  Ratra \& Peebles (1988) \\
& \\
$V_0\exp{(\lambda\phi^2)}/\phi^\alpha, \alpha > 0$ & Brax \& Martin (1999,2000)\\
& \\
$V_0(\cosh{\lambda\phi} - 1)^p$, & Sahni \& Wang (2000)\\
& \\
$V_0 \sinh^{-\alpha}{(\lambda\phi)}$, & Sahni \& Starobinsky (2000),
Ure\~{n}a-L\'{o}pez \& Matos (2000)\\
& \\
$V_0(e^{\alpha\kappa\phi} + e^{\beta\kappa\phi})$ & Barreiro, Copeland \& Nunes (
2000)\\
& \\
$V_0(\exp{M_p/\phi} - 1)$, & Zlatev, Wang \& Steinhardt (1999)\\
& \\
$V_0[(\phi - B)^\alpha + A]e^{-\lambda\phi}$, & Albrecht \& Skordis (2000)\\

& \\
\hline
\end{tabular}
\caption{}
\end{minipage}
\end{center}
\end{table*}

Several of the quintessential potentials listed in table 1
have been inspired by field theoretic ideas
including supersymmetric gauge theories and supergravity, 
pseudo-goldstone boson models, etc.
However accelerated expansion can also arise in models
with: (i) topological defects such as a frustrated network of cosmic 
strings ($w \simeq -1/3$) and domain walls ($w \simeq -2/3$)\cite{spergel98};
(ii) scalar field lagrangians with non-linear kinetic terms and no potential 
term (k-essence \cite{mukhanov}); (iii) vacuum polarization associated with an 
ultra-light scalar field \cite{sahni98,parker99};
(iv) non-minimally coupled scalar fields \cite{nonmin}; (v) fields that couple to matter
\cite{amendola}; (vi) scalar-tensor theories of gravity \cite{scalartensor};
(vii) brane-world models \cite{maartens,copeland,lidsey,sami,dvali,ss02} etc.

Scalar field based quintessence models can be broadly divided into two classes:
(i) those for which $\phi/\mpl \ll 1$ as $t \to t_0$, 
(ii) those for which 
$\phi/\mpl \geq 1$ as $t \to t_0$ ($t_0$ is the present time).
An important issue concerning the second class of models
is whether quantum corrections become important
when $\phi/\mpl \geq 1$ and their possible effect on the quintessence
potential \cite{kolda98}. 
One can also ask whether a given
choice of parameter values is `natural'. Consider 
for instance the potential
$V = M^{4+\alpha}/\phi^\alpha$, current observations 
indicate
$V_0 \simeq 10^{-47}$GeV$^4$ and $\alpha \lleq 2$, which together suggest
$M \lleq 0.1$ GeV 
(smaller values of $M$ arise for smaller $\alpha$)
it is not clear whether such small parameter values can be motivated 
by current models of high energy physics.

Finally, it would be enormously interesting if one and the same
field could give rise to both Inflation and dark energy. Such models
have been discussed both in the context of standard inflation \cite{peebles99}
and brane-world inflation \cite{copeland,lidsey,sami}, we briefly
discuss the second possibility below. 

\subsection{Quintessential Inflation in Braneworld scenario's}

In the 4+1 dimensional brane scenario
inspired by the Randall-Sundrum \cite{rs} model, matter fields are confined
to a three dimensional `brane' which is embedded in a four dimensional 
`bulk' geometry.
The equation of motion of a scalar field propogating on the brane is
\beq
{\ddot \phi} + 3H {\dot \phi} + V'(\phi) = 0.
\label{eq:kg}
\eeq
where \cite{brane}
\beq
H^2 = \frac{8\pi}{3 M_4^2}\rho (1 + \frac{\rho}{2\lambda_b}),\;
\rho = \half{\dot\phi}^2 + V(\phi),
\label{eq:brane}
\eeq
and $\lambda_b$ is the brane tension.
The additional term $\rho^2/\lambda_b$ in (\ref{eq:brane}) arises due to
junction conditions imposed at the bulk-brane boundary. The presence of
this term {\em increases
the damping} experienced by the scalar field as it rolls down its
potential. This effect is reflected in the slow-roll parameters,
which in braneworld models 
(for $V/\lambda_b \gg 1$) have the form \cite{maartens}
\beq
\epsilon \simeq 4\epsilon_{\rm FRW} (V/\lambda_b)^{-1},\,
\eta \simeq 2\eta_{\rm FRW} (V/\lambda_b)^{-1}.
\label{eq:slow-roll}
\eeq
Clearly slow-roll ($\epsilon, \eta \ll 1$) is easier to achieve when
$V/\lambda_b \gg 1$ and on this basis one can expect inflation
to occur even for the very steep potentials associated with 
quintessence models including
$V \propto e^{\lambda\phi}$, $V \propto \phi^{-\alpha}$ etc.
Inflation in these models has been extensively discussed in 
\cite{copeland,lidsey,sami,majumdar} within the framework of a scenario in which
reheating takes place unconventionally,
through inflationary particle production. This leads to
an enormous difference between the 
energy in the inflaton and in radiation at the end of inflation:
$\rho_\phi/\rho_{rad}\vert_{\rm end} \sim 10^{16}$. 
Since the potential driving inflation 
is steep, the
post-inflationary expansion in these models is driven by the kinetic
energy of the scalar field, so that $w_\phi \simeq 1$,
$\rho_\phi \propto a^{-6}$ and $a \propto t^{1/3}$.   
(Because radiation decreases at the slower rate $\rho_{\rm rad} \propto a^{-4}$
the scale factor changes to $a \propto t^{1/2}$
after the density in the inflaton and in radiation equalize. 
This usually takes place at a low temperature $T_{\rm eq} \sim {\rm few}$ GeV.)

As demonstrated in \cite{copeland,lidsey,sami,majumdar}
inflation can occur for several of the quintessence potentials 
discussed in the previous section
but for a rather narrow region of parameter
space (see figure 2). It also appears that quintessential inflation
generates a large gravity wave background which could
be in conflict with big bang nucleosynthesis considerations \cite{sami}.

\begin{figure}[tbh!]
\centerline{
\psfig{file=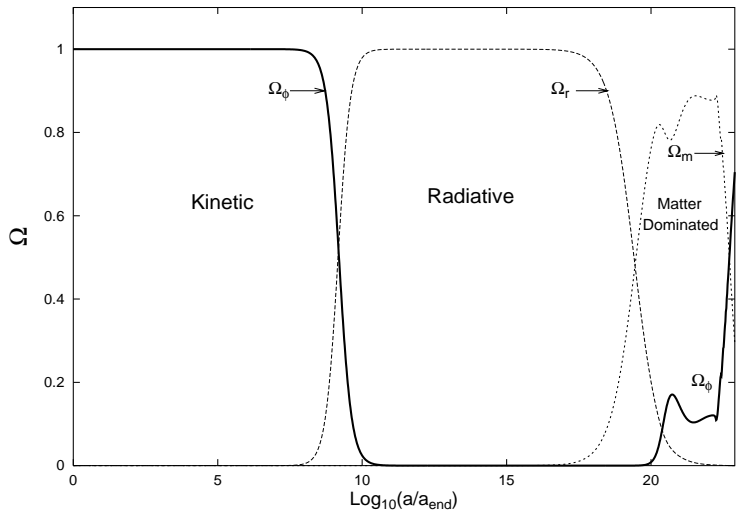,width=10cm}}
\caption{\footnotesize
The post-inflationary  density parameter $\Omega$ is plotted for
the scalar field (solid line) radiation (dashed line) and cold dark
matter (dotted line)
in the quintessential-inflationary model
decribed by (\ref{eq:cosh}) with $p = 0.2$.
Late time oscillations of the scalar field ensure that
the mean equation of state turns negative
$\langle w_\phi\rangle \simeq -2/3$, giving rise to the current
epoch of cosmic acceleration with $a(t) \propto t^2$ and present day values
$\Omega_{0\phi} \simeq 0.7, \Omega_{0m} \simeq 0.3$.
From Sahni, Sami and Souradeep \cite{sami}.}
\bigskip
\medskip
\label{fig:sami}
\end{figure}

\subsection{Reconstructing the cosmic equation of state}
Although fundamental theories such as Supergravity or M-theory do provide a
number of possible candidates for quintessence they do
not uniquely predict its potential $V(\phi)$. Therefore it becomes meaningful
to reconstruct $V(\phi)$ and the cosmic equation of state $w = P/\rho$
directly from observations in a {\em model independent} manner
\cite{saini,recon,recon1,recon2}.
This is possible to do if one notices that the scalar field potential
as well as its equation of state can be directly expressed in terms
of the Hubble parameter and its derivative
\ber
{8\pi G\over 3H_0^2} V(x)\ &=& {H^2\over H_0^2}
-{x\over 6H_0^2}{dH^2\over dx} -{1\over 2}\Omega_m\,x^3,
\label{eqn:Vzed}\\
{8\pi G\over 3H_0^2}\left({d\phi\over dx}\right)^2 &=&
       {2\over 3H_0^2 x}{d\ln H\over dx}
  -{\Omega_m x\over H^2}, ~ x = 1+z,\\
&&\\
w_\phi (x) \equiv {p\over\varepsilon} &=&
\frac{(2x/3) d\ln H/dx -1}{1-\left(H^2_0/H^2\right)
\Omega_m x^3}.
\eer
Since the Hubble parameter is related to the luminosity distance
\beq
H(z) = \left[ \frac{d}{dz} \left( \frac{d_L(z)}{1+z} \right) \right]^{-1},
\eeq
one can determine both the quintessence potential $V(\phi)$ as well as
reconstruct its equation of state $w_\phi (z)$ provided 
the luminosity  distance $d_L(z)$ is 
known from observations. A three parameter ansatz for estimating the 
luminosity  distance was proposed in \cite{saini}. Results from that paper 
reproduced in
figure 3 indicate that only a small amount of evolution in $w_\phi (z)$
is permitted by current SnIa observations. The presence of 
a cosmological constant
is therefore in good agreement with these results.

\begin{figure}[tbh!]
\centerline{
\psfig{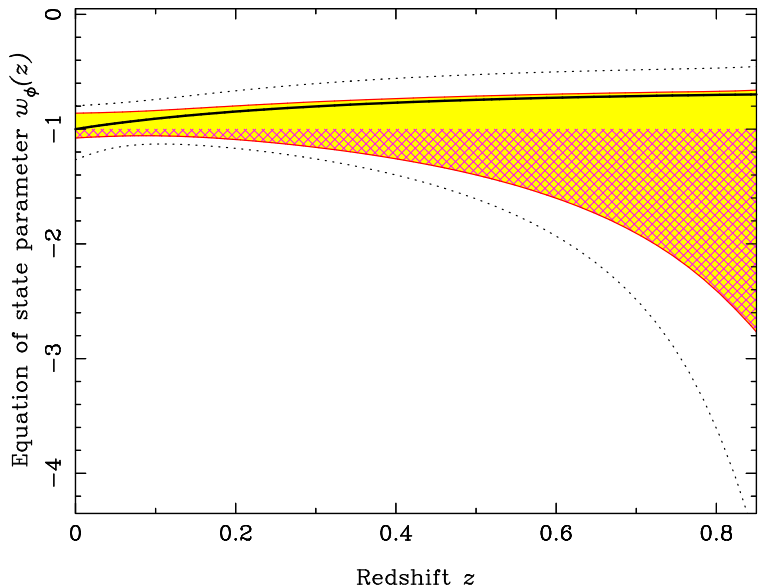}}
\caption{\footnotesize
The equation of state of dark energy/quintessence is 
reconstructed from observations
of Type Ia high redshift supernovae in a model independent manner. 
The equation of state
satisfies $-1 \leq w_\phi \leq -0.8$ at $z = 0$;
and
$-1 \leq w_\phi \leq -0.46$ at $z = 0.83$ ($90\%$ CL),
$\Omega_m = 0.3$ is assumed.
From Saini, Raychaudhury, Sahni and Starobinsky \cite{saini}. }
\bigskip
\medskip
\label{fig:state}
\end{figure}

\begin{figure}[tbh!]
\centerline{
\psfig{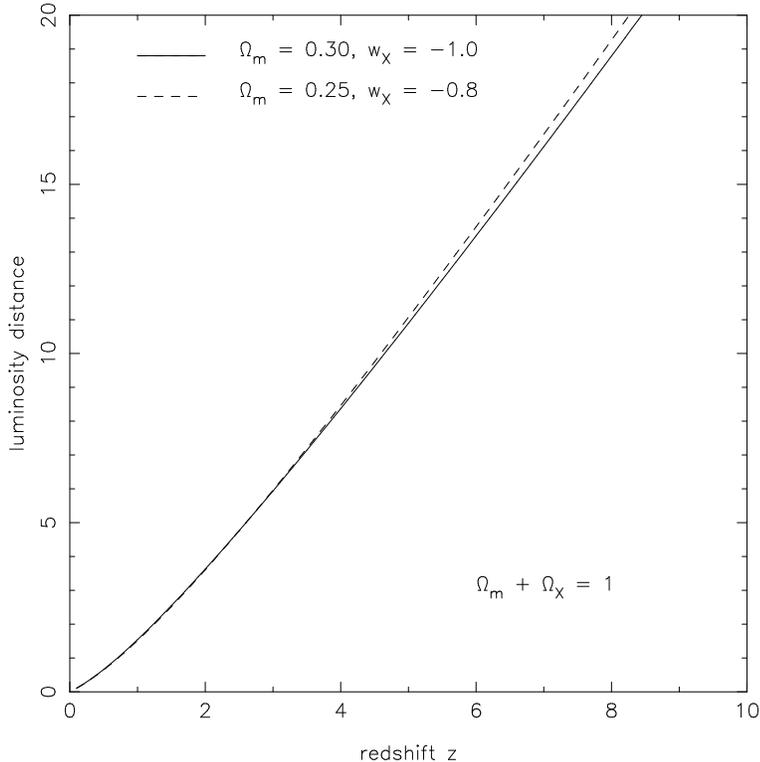}}
\caption{\footnotesize
The near degeneracy in the luminosity distance is shown for the
pair of cosmological models with $\lbrace \Omega_m = 0.3, w_X = -1.0\rbrace$
and $\lbrace \Omega_m = 0.25, w_X = -0.8\rbrace$.
}
\bigskip
\medskip
\label{fig:degen}
\end{figure}

A word of caution should be added: as
shown in figure 4 a near degeneracy exists between the equation of state 
of dark energy
and the value of $\Omega_m$. 
The latter should therefore be known to better than $5\%$ accuracy for
the reconstruction program to yield very accurate results 
(see also \cite{recon1}).

\subsection{Probing dark energy using the Statefinder statistic}

An issue of the utmost importance is whether dark energy 
(equivalently quintessence) 
is a cosmological constant or whether it has a fundamentally different
origin. A new dimensionless statistic `Statefinder',
recently introduced by Sahni, Saini, Starobinsky and Alam \cite{sss01} has the
power to discriminate between different forms of dark energy and may
therefore be a good diagnostic of cosmological models.

The Statefinder pair $\lbrace r,s\rbrace$ is constructed from the
scale factor of the universe and its derivatives and probes
the cosmic equation of state and its rate of change.
It extends 
the hierarchy of geometrical cosmological parameters to four:
$\lbrace H,q,r,s \rbrace$, 
where  $H = ({\dot a}/a)$, $q = - H^{-2}({\ddot a}/a)$,
\ber
r &=& \frac{\stackrel{...}{a}}{a H^3} =
1 + \frac{9 w}{2}\Omega_\phi (1+w)
- \frac{3}{2}\Omega_\phi \frac{\dot w}{H},\nonumber\\
s &=& \frac{r - 1}{3(q - 1/2)}
= 1 + w - \frac{1}{3}\frac{\dot w}{w H}.
\eer

From figure 5 we see that while $r$ {\em remains fixed} at $r = 1$
in a universe containing matter and a cosmological constant,
the value of $r$ decreases steadily for time varying 
forms of dark energy.
The Statefinder statistic can therefore help differentiate
a cosmological constant from:
(i) dark energy with a {\em time-independent} equation of state 
(referred to in \cite{sss01} as Quiessence) and (ii) dark energy with a 
time-dependent equation of state 
(referred to in \cite{sss01} as Kinessence).

\begin{figure}[tbh!]
\centerline{
\psfig{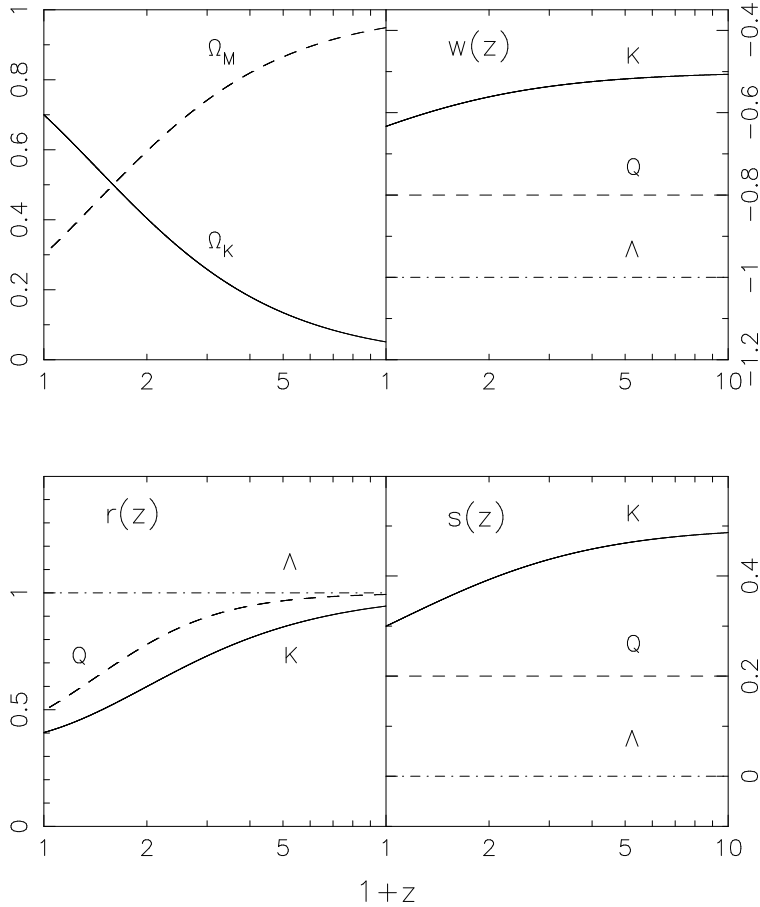} }
\caption{\footnotesize
The Statefinder pair $\lbrace r,s\rbrace$ is shown
for dark energy consisting of a cosmological constant $\l$,
Quiessence `Q' with an unevolving equation of state $w = -0.8$ and
the inverse power law tracker model $V= V_0/\phi^2$,
referred to as Kinessence `K'.
The lower left panel shows $r(z)$ while the lower right panel shows $s(z)$.
Kinessence has a time-dependent equation of state which is shown in
the top right panel. 
The fractional density in matter and
Kinessence is shown in the top left panel.
The ability of the Statefinder pair $\lbrace r,s\rbrace$
to differentiate between the different forms of dark energy is amply
demonstrated by this figure which is reproduced from
Sahni, Saini, Starobinsky and Alam \cite{sss01}.
}
\bigskip
\medskip
\label{fig:statefinder}
\end{figure}

\section{Horizons in a $\l$-dominated universe}

The conventional viewpoint regarding the future evolution of 
a matter-dominated universe can be summarised by the following pair of
statements:

[A] If the universe is spatially open or flat then it will expand
for ever.

[B] Alternatively, if the universe is spatially closed then its
expansion will be followed by recollapse.

In a $\l$-dominated universe [A] and [B] are replaced by

[A'] A spatially open or flat universe ($\kappa = 0, -1$)
will recollapse if the $\l$-term is
constant and negative.

[B'] A closed universe ($\kappa = +1$)
with a constant positive $\l$-term can, under certain
circumstances, expand forever.

Recent CMB observations indicate that our universe is close to
being spatially flat, therefore, if we wait long enough we will find that
the expansion of our universe rapidly approaches the
de Sitter value $H = H_\infty = \sqrt{\l/3} = H_0\sqrt{1 - \Omega_m}$, while
the density of matter asymptotically declines to zero $\rho_m \propto a^{-3}
\to 0$.

Density perturbations in such a universe will freeze to a constant value
if they are still in the linear regime,
but the acceleration of
the universe will not affect gravitationally bound systems on present scales
of $R < 10$h$^{-1}$ Mpc (which includes our own galaxy as well as galaxy
clusters).
The universe at late times will therefore consist of islands of matter
immersed in an accelerating sea of dark energy: `$\l$'.

In such a universe the local neighborhood
of an observer from which he/she is able to receive signals will eventually
contract and shrink so that even those regions of the universe which
are currently observable to us will eventually be hidden from view.
As an illustration consider an event at ($r_1,t_1$) which we wish to observe
at our location at $r = 0$, then
\beq
\int_0^{r_1}\frac{dr}{\sqrt{1-\kappa r^2}} = \int_{t_1}^{t}\frac{dt'}{a(t')}.
\label{eq:future1}
\eeq
An observer at $r=0$ will be able to receive signals from any event
(after a suitably long wait) provided 
the integral in the RHS of (\ref{eq:future1}) diverges (as $t \to \infty$).
For $a \propto t^p$,
this implies $p < 0$, or a decelerating universe.
In an accelerating universe the integral converges, signaling the presence
of an event horizon.
As a result one can only receive signals from those events which satisfy
\beq
\int_0^{r_1}\frac{dr}{\sqrt{1-\kappa r^2}} \leq 
\int_{t_1}^{\infty}\frac{dt'}{a(t')}.
\label{eq:future2}
\eeq
For de Sitter-like expansion
$a = a_1\exp{H(t - t_1)}, H = \sqrt{\l/3}$, we get $r_1 \equiv r_H$
and
$R = a_1r_H = H^{-1}$
where $R$ is the proper distance to the event horizon.
In such a universe light emitted by distant objects gets redshifted and 
declines in intensity 
(an analogous situation occurs for an object falling through the
horizon of a black hole.)
As a result
comoving observers once visible, gradually disappear
from view as the universe accelerates under the influence of $\l$. 
One consequence of this interesting phenomenon is that
at any given instant of time, $t_0$, one can
determine a redshift $z_H$, which will define for us 
the `sphere of influence' of our civilization. 
Celestial objects with $z > z_{H}$ will always remain
inaccessible to signals emitted by our civilization at $t \geq t_0$.
For a $\l$-dominated universe with $\Omega_m = 0.3$,
$\Omega_\l = 0.7$ one finds $z_{H} = 1.80$ \cite{ss99}.
More generally,
horizon's exist in a universe which begins to perpetually accelerate
after a given point of time \cite{hellerman01,fischler01}. 
(To this general category belong models of dark energy 
with equation of state $-1 < w < -1/3$, as well as `runaway
scalar fields' \cite{stein01} which satisfy 
$V,V',V'' \to 0$ and $V'/V, V''/V \to 0$ as 
$\phi \to \infty$.)
Since the conventional S-matrix approach may not work in a universe with an
event horizon, such a cosmological model
may pose a serious challenge to a fundamental theory of
interactions such as string theory. 
Possible ways of cutting short `eternal acceleration' (thereby 
avoiding horizons) involve scalar fields with non-monotonic potentials. 
For instance a flat potential with a local minimum 
will have a negative equation of state during slow roll,
which will increase to non-negative values after the cessation of
slow roll and the commencement of oscillations (provided the potential
is sufficiently steep in the neighborhood of the minimum).
An example is provided by a massive scalar field $V(\phi) = m^2\phi^2/2$
for which the epoch of accelerated expansion is a transient.
Other possibilities are discussed in \cite{fischler01,kolda01,dvali,ss02,cline01,li02}.

\section{Conclusions}
One of the major concerns of  cosmology today is the 
nature of dark energy (quintessence).
While a cosmological constant appears to be the simplest option, formidable
fine-tuning problems which confront the latter have led to theoretical
models being developed
in which both the dark energy as well as its equation of state are
functions of time. 
Type Ia supernovae currently provide the strongest evidence for
dark energy. The very recent observations of a supernova at $z \simeq 1.7$
appear to rule out simple extinction and evolution effects as major
causes for the diminishing light flux from these high-$z$ objects. It therefore
appears that the dark energy is `real' and that the universe was decelerating
at $z > 0.5$ \cite{turner01}.

The observational situation is likely to improve as results from both deep and
extensive galaxy and galaxy cluster survey's come in. It is well known that the
presence of a cosmological constant 
changes the shape of the two point galaxy-galaxy correlation
function by increasing the strength of clustering
on large scales \cite{esm90}.
Additionally, since $\l$ slows down the growth of gravitational
clustering, galaxy clusters are expected to be more abundant at high redshift
in $\l$CDM than in the standard cold dark matter scenario (SCDM).
(Tracker fields can give rise to a large smooth component
of matter at high $z$, as a result
gravitational clustering takes 
place at a slower rate 
in tracker-quintessence models than in $\l$CDM \cite{bern01}.)
Both these effects are likely to be tested in the near future \cite{holder01}.
Indeed, recent results which measure the gravitational clustering of over 100,000
galaxies in the 2dF survey,
determine a matter 
power spectrum which is consistent with $\l$CDM and inconsistent
with SCDM \cite{peacock01}.

Combined results from CMB probes (MAP, PLANCK), galaxy surveys (2dF, SDSS, DEEP),
weak lensing statistics and the possibility of a dedicated supernova
telescope (SNAP) give rise to expectations that the nature of
dark energy will be understood (at least at the phenomenological level)
in the not too distant future.


\section*{References}

\end{document}